\documentclass[aps,prd,
nofootinbib,
preprintnumbers,
twocolumn,
showpacs, 
floatfix,
superscriptaddress
]{revtex4}

\def\mnras{Mon. Not. R. Astron. Soc.}

\begin{document}

\title{A comment on ``Exclusion of the remaining mass window for primordial
  black holes...'', arXiv:1401.3025}

\author{Fabio Capela}
\affiliation{Service de Physique Th\'{e}orique, 
Universit\'{e} Libre de Bruxelles (ULB),\\CP225 Boulevard du 
Triomphe, B-1050 Bruxelles, Belgium}

\author{Maxim Pshirkov} \affiliation{Sternberg Astronomical Institute,
  Lomonosov Moscow State University, Universitetsky prospekt 13, 119992,
  Moscow, Russia}

\affiliation{Pushchino Radio Astronomy Observatory, 
Astro Space Center, Lebedev Physical Institute  Russian Academy of Sciences,  
142290 Pushchino, Russia}

\affiliation{Institute for Nuclear Research of the Russian Academy of
  Sciences, 117312, Moscow, Russia}

\author{Peter Tinyakov}
\affiliation{Service de Physique Th\'{e}orique, 
Universit\'{e} Libre de Bruxelles (ULB),\\CP225 Boulevard du 
Triomphe, B-1050 Bruxelles, Belgium}

\maketitle

In the recent paper \cite{Pani:2014rca}, A.~Loeb and P.~Pani have proposed a
new energy loss mechanism by primordial black holes (PBH) of the mass $10^{17}
- 10^{26}$~g passing through a neutron star (NS). This mechanism is claimed to
be many orders of magnitude more efficient than the dynamical friction
mechanism that was used in the previous estimates \cite{Capela:2013yf}. As a
result, the constraints on the abundance of PBH were improved by many orders
of magnitude.  In this comment we point out a potential problem in the
calculations of Ref.~\cite{Pani:2014rca} that may invalidate their result. In
that case, the new mechanism would give parametrically the same energy loss as
the dynamical friction, and would not imply a significant improvement of the
constraints on PBHs.

The difference between the two calculations of the energy losses is that in
Ref.~\cite{Capela:2013yf} the NS matter is treated as approximately
pressureless during its interaction with the PBH. In this approach, the energy
losses result from gravitational scattering of individual neutrons on the
passing PBH. On the contrary, in Ref.~\cite{Pani:2014rca} the NS matter is
treated as a gravitating fluid, and the energy losses result from excitation
of acoustic oscillations of this fluid by a moving PBH.

Despite apparent difference in the approaches, the huge mismatch in the
results is difficult to believe because of causality reasons.  A PBH falling
on a NS attains a relativistic velocity $\sim 0.6c$ (in the middle of the
star). This velocity is larger (in the center) or much larger (in the outer
core and in the crust) than the speed of sound. When the NS matter gets
perturbed by the passing PBH, there is no time for it to organize itself in
any kind of acoustic waves because of the slow sound speed. In other words, at
the time scale of the interaction with the PBH the NS matter behaves
approximately as pressureless. In this approximation, the answer for the
energy loss has been calculated by Chandrasekhar \cite{Chandrasekhar}, the
corresponding mechanism being known as the dynamical friction. Within this
picture, the alternative approach in terms of the sound waves may be viewed
simply as a different way of assessing the energy losses in which the
perturbation created by the BH, with the energy losses already encoded in it,
is first represented as a sum of acoustic modes, and then energies of the
individual modes are summed up. Thus, in the limit when the PBH velocity is
much larger than the speed of sound, the answer should be the same whatever
approach is used.

The fact that the pressureless approximation is accurate in the case of a
supersonic motion has been confirmed by an explicit calculation in
Ref.~\cite{Ostriker}, where it was shown that already for a PBH passing
through a homogeneous medium with velocity larger than $2-3$ times the sound
speed, the energy loss to the emission of sound waves reproduces exactly the
dynamical friction formula. Thus, no large differences between the acoustic
oscillation approach and the dynamical friction one should be expected in the
case of a PBH crossing a NS.

With this understanding in mind, consider how the large enhancement factor of
Ref.~\cite{Pani:2014rca} comes about. The approach of Ref.~\cite{Pani:2014rca}
is based on the calculation of the excitations of seismic modes of a NS by a
passing PBH. The dominant contribution was argued to come from the fundamental
$n=0$ modes, the so-called $f$-modes that are, loosely speaking, surface waves
at the surface of the NS core. These waves are supported by the combination of
gravity and pressure, not the pressure alone as usual sound waves, hence a
possible difference with the calculation of Ref.~\cite{Ostriker}. Note
however, that these waves still do not propagate faster than the speed of
sound, so the dynamical friction result is still expected in the supersonic
case. The contributions of these $f$-modes to the total energy loss were
calculated numerically up to $l=100$ and found to follow the behavior
\begin{equation}
E_l \propto {m_{\rm BH}^2\over R} {1\over l^n},
\label{eq:El}
\end{equation}
where $m_{\rm BH}$ is the BH mass, $R=10$km is the NS radius and $n\sim 0.5$
is related to the equation of 
state (EOS) in the NS core. The coefficient $m_{\rm
  BH}^2/ R$ is, up to a numerical factor, the energy loss due to the dynamical
friction \cite{Capela:2013yf}. So, the sum $\sum_l l^{-n}$ represents the
enhancement compared to the dynamical friction result (again, up to a
numerical factor). The sum is divergent. The authors of
Ref.~\cite{Pani:2014rca} cut the sum at the BH size which corresponds to
$l_{\rm max} \sim R M_{\rm Pl}^2/m_{\rm BH}\sim 10^{10}$ for $m_{\rm BH}
=10^{24}$~g. Extrapolating eq.~(\ref{eq:El}) from $l=100$ to $l=10^{10}$ and 
evaluating the sum then gives the enhancement factor $l_{\rm
  max}^{1-n}\sim 10^5 \gg 1$.

We think that the discrepancy between this result and the dynamical friction
calculation may be explained by the breaking of eq.~(\ref{eq:El}) beyond $l$
of order several hundred. The physical reason for this is that the core-crust
transition in a NS --- the region relevant for deriving eq.~(\ref{eq:El}) ---
has a finite thickness (cf.~\cite{interfacial-modes} where a similar idea is
discussed in the context of Solar oscillations). The characteristic distance
scale can be estimated as $d=\rho/(d\rho/dr)$. From the realistic NS density
profile \cite{profile} we have obtained $d\sim 30$~m, which corresponds to
$l_d\sim 500$.  At the wavelengths much smaller than $d$ (equivalently, for
$l \gg l_d$) the character of the $f$-modes must change and eq.~(\ref{eq:El})
should get modified. Note that eq.~(\ref{eq:El}) has to be extrapolated to
much larger values $l\sim 10^{10}$ in order to get the claimed enhancement
factor.

In mathematical terms, the break in $E_l$ can be understood as
follows. Equation for the acoustic perturbations in the gravitating fluid, 
in the Cowling approximation and in the limit of vanishing buoyancy frequency 
(which is identically zero for a polytropic EOS), 
can be written as \cite{f-modes}
\begin{equation}
\xi'' + \left( {4\over r} + {\rho'\over \rho}\right) \xi' 
+ \left({\rho'\over r\rho} - {l(l+1)-2\over r^2} \right)\xi 
= -{\omega^2\over v_s^2} \xi,
\label{eq:2}
\end{equation}
where the prime denotes the derivative with respect to $r$, $\xi$ is the
angular displacement, $v_s$ is the sound speed and $\rho$ is the
density. Note the appearance of the combination $\rho'/\rho$, the inverse
distance scale. 
To understand the behavior of solutions one may eliminate the
first derivative by an appropriate change of variables $\xi= f(r) \eta$. This
gives
\begin{equation}
\eta'' + \left( - {\rho''\over 2\rho} - {\rho'\over r \rho} + {1\over 4} 
{\rho'^2\over \rho^2} - {l(l+1)\over r^2}
\right)\eta = -{\omega^2\over v_s^2} \eta.
\label{eq:}
\end{equation}
Since $\rho'/\rho=-g/v_s^2$, $g$ being the gravitational acceleration, the
$\rho$-dependent terms grow toward the surface where the sound speed becomes
small. For a polytropic EOS, neglecting the variation of $g$, one has
$\rho'/\rho \propto -(R-r)^{-1}$, $R$ being the boundary of the star.  As one
can see from eq.~(\ref{eq:}), in the region away from the surface and for
large $l$, the term $l(l+1)/r^2$ dominates and there are no solutions except
for normal sound waves with the dispersion relation $\omega^2 = v_s^2
(k_r^2+k_{||}^2)$, where $k_r$ and $k_{||}$ are the radial and angular
components of the momentum. The surface waves, for which $k_r\sim 0$ and the
dispersion relation is linear in $k_{||}$, have no support in this
region. They are concentrated in the layer of the thickness $\sim R/l$ next to
the boundary where the $l(l+1)/r^2$ term can be balanced by other
contributions. This is a known result \cite{f-modes}.

In a real NS, the perturbation equations are more complicated
\cite{Ruoff:2000gc}. Still, we think that eq.~(\ref{eq:2}) grasps
the main features and is sufficient for a rough estimate. For a realistic NS
density profile, the $\rho$-dependent terms in eq.~(\ref{eq:}) grow with
increasing $r$ and reach the value of $\sim 1/d^2$ around the core-crust
transition region, as set by the finite thickness $\sim d$ of this region, and
then grow to much larger values toward the surface of the star. Thus, for
$l\lesssim l_d\sim R/d$, eq.~(\ref{eq:}) allows for the existence of the
$f$-modes which peak in the core-crust transition region and may produce
eq.~(\ref{eq:El}). However, at $l\gtrsim l_d$ the term $l(l+1)/r^2$ in
eq.~(\ref{eq:}) becomes dominant in this region. Therefore, the $f$-modes with
such $l$ peak in the crust and are exponentially small around the core-crust
transition. For this reason they know nothing about the EOS in the core, and
thus the corresponding $E_l$ cannot follow eq.~(\ref{eq:El}) with $n\simeq
0.5$ characteristic of the core.  The behavior of $E_l$ must, therefore,
change.

To summarize, the extrapolation of eq.~(\ref{eq:El}) from $l=100$ to $l\sim
10^{10}$ (for $m_{\rm BH}=10^{24}$~g) is unjustified. Moreover, one should
expect a suppression in eq.~(\ref{eq:El}) at $l$ of order several hundred. If
only modes up to $l\sim 100$ in eq.~(\ref{eq:El}) are summed up, the answer is
not different from the dynamical friction result apart from possibly a
numerical factor of order 1, in accord with the causality arguments.

The expected break in $E_l$ may not be the only problem of the calculation of
Ref.~\cite{Pani:2014rca}, the other one being the justification of the linear
approximation. The justification given in Ref.~\cite{Pani:2014rca} --- that
the energy losses are much smaller than the rest mass of the PBH --- does not
sound convincing, because it is not clear how does this fact translate into
smallness of the non-linear terms in the full hydrodynamic equations --- the
terms that have been omitted when deriving the linearized mode equations for
perturbations. Certainly, a more detailed study of this issue is required.

We thank G.~Rubtsov and M.~Tytgat for helpful discussions,
and P.~Pani and A.~Loeb for useful references and patience in explaining their
point of view, with which we finally do not agree.

\end{document}